# Implementation of an Efficient RBAC Technique of Cloud Computing in .Net Environment

**Ruhi Gupta**
**Department Of Computer Science, Punjabi University, Patiala, India**

*Abstract:*
*Cloud Computing is flourishing day by day and it will continue in developing phase until computers and internet era is in existence. While dealing with cloud computing, a number of security and traffic related issues are confronted. Load Balancing is one of the answers to these issues. RBAC deals with such an answer. The proposed technique involves the hybrid of FCFS with RBAC technique. RBAC will assign roles to the clients and clients with a particular role can only access the particular document. Hence identity management and access management are fully implemented using this technique.*

*Keywords: ABAC, Cloud Computing, IBAC, FCFS, RBAC*

## I. Introduction:

Cloud Computing is a platform which aims to provide shared data to its clients at the same time. It is amongst the buzzwords in today's era. Whether we open an IT magazine or open any website, cloud computing concept is everywhere. As the count of clients for the access of same data increases, catastrophe may occur. Cloud Computing offers various service models. It can be software as a service model, offering softwares on a single platform. It can be platform as a service model which offers a platform from where the softwares and data can be accessed. Or it can be infrastructure as a service which provides the security and backup services. Load balancing concept cannot lag behind while dealing with cloud computing. Load balancing can be implemented during runtime or can be predefined. When predefined, all he nodes of the network can have a fix number of load on them. But during runtime, the overloaded node can transfer its over load to the underloaded node so as to make it a balanced node. RBAC is one of the techniques of the load balancing during runtime.

## II. RBAC in Load Balancing:

Concept of RBAC is the most captivating concept among all load balancing techniques. Through RBAC, only a limited number of clients can access the particular document thus decreasing the traffic load over a single network and increasing the security of the clouds. Restriction is put on the roles of the users, hence leading to the security of the cloud. There are various types by which the load balancing can be subdivided among which are static and dynamic techniques. RBAC is a kind of static one but along with FCFS, dynamic roles are being set for them. Access control models restrict or enable the access to any data. There are various access control models i.e. MAC (Mandatory access control model), DAC (Discretionary access control model) and RBAC (role based access control model). MAC is responsible for assigning roles to the users. DAC is responsible for checking the authorisation of the users. So RBAC is a combination of MAC and DAC. RBAC can be implemented whether by using networking environment or .NET environment. The main focus is to set priority according to the roles of the various users. If one door is closed, then second door opens for its access. Hence RBAC plays an important role in Networking systems.

RBAC consists of two phases:

1) One or more roles are given to the clients.
2) Roles are the cross checked.

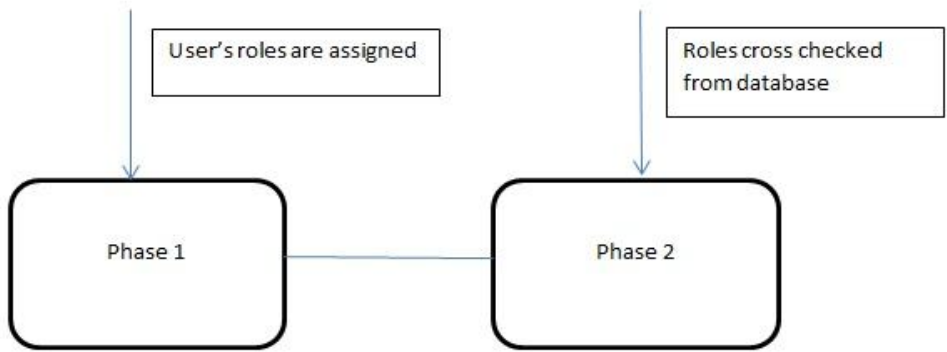

RBAC has three components: Core RBAC, Hierarchal RBAC and Constrained RBAC.

In Core RBAC, the user's login to the system and operations are performed by them according to their roles. In hierarchal RBAC, one client can always depend on the existence of the other. For e.g. in any company, employee can get permission only if granted by manager. This also works as a constraint RBAC.

A feature of RBAC involves:

1. It provides the security of the data by limiting the roles to the users.
2. It reduces the transaction time of the data by limiting the users.
3. Backup of data can be taken.

### III. Related Work:
Till now various researchers had developed various enhanced models of RBAC. Geetanjli et. al. [3] in the research work proposed a technique in which it was easy to migrate the tasks from one system to another. XML implementation Backup and Restoration policy was applied to have backup of the data.

In this way, RBAC was made more enhanced. Parminder et. al. [1] in their research work implemented enhanced RBAC in .NET environment and windows azure for database. was done in their research work. Restriction policy was applied to have more secured system. Along with the backup of data, restriction was applied on the number of transactions which made the data more secure from hackers.

Er. Amandeep kaur et. al. [2] discovered a Cross Breed algorithm which involves the hybrid of FCFS, RBAC and a priority algorithm. Jason and Koch had given the main concept of RBAC by introducing the admin and the user's concept by which the load can be balanced.

### IV. Implementation:
The proposed work involves the implementation of the RBAC with FCFS in .NET environment so as to have an easy and secure access to cloud. Users first make their account in the organisation. Then RBAC is implemented i.e. roles are assigned and authorised to the users. After that, the FCFS technique is applied. This whole working is shown by the flowchart Fig. 1 below:

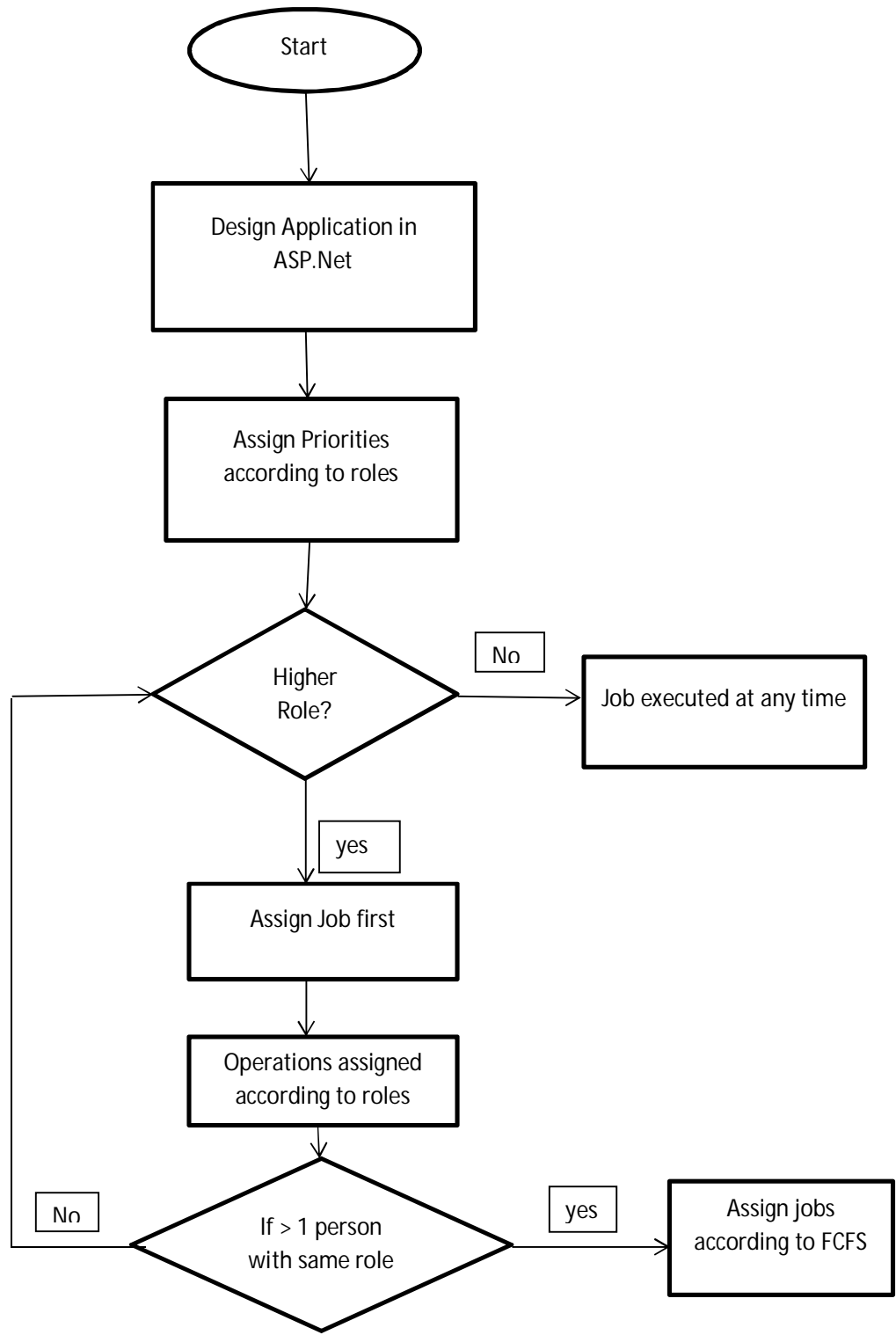

*Fig.1 : Working of the technique*

According to this flowchart which depicts the proposed work, clients first login their accounts according to the roles they have in the organisation. For e.g. in the application developed in .NET environment, roles are assigned in order of:

1. Admin
2. President
3. GM
4. Manager       Priority decreases
5. Executive

The figure 2 shows the login page of .NET.

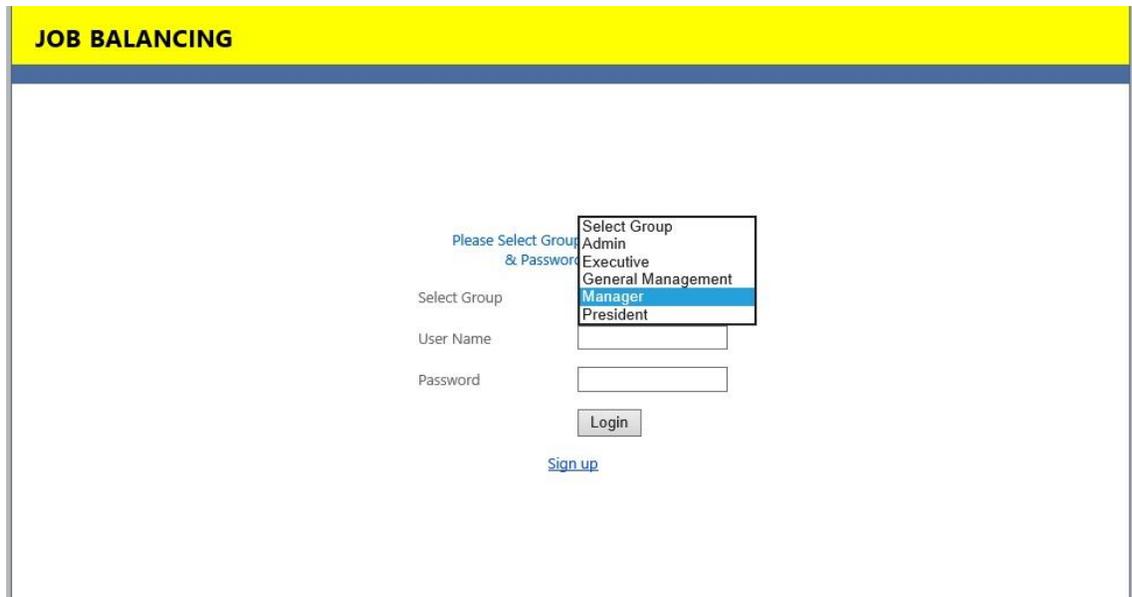

*Fig. 2: Login Page*

After clicking on the *create an account* on the home page, the attributes have great role. All the attributes once submitted are stored in the database i.e. Microsoft SQL Server 2008 in this work. Attributes works in following phases:

a) New Client: Every new client is assigned the new id containing the same list of attributes. It is termed as the requestor who wants to have access to the cloud.
b) Job: Every client who wants to access the job that is assigned the job after checking the correct identity of the client. It is termed as the service requested by the client from the cloud.
c) Conditions: It is the date and time where a particular job may or may not be available. It is not listed in the attribute category in practice.
RBAC is a combination of attribute based access control i.e. ABAC and the identity based access control i.e. IBAC. These both lead to the security and saving of the time and money.

After having the account, and login to the page, it is found that the admin manages as well as assigns the jobs to the various users of the organisation. It is shown by figure 3.

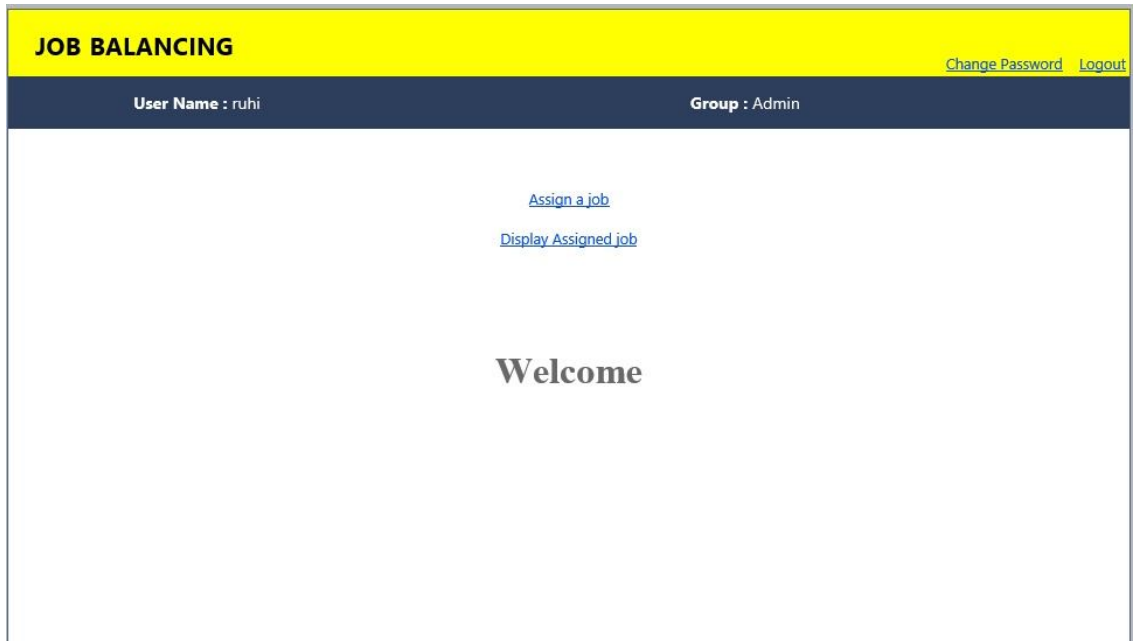

*Fig. 3: Admin manages the jobs*

Whereas all the clients of lesser roles can only access the jobs listed by the admin. So according to the roles, . priorities are assigned. It is shown in figure4.

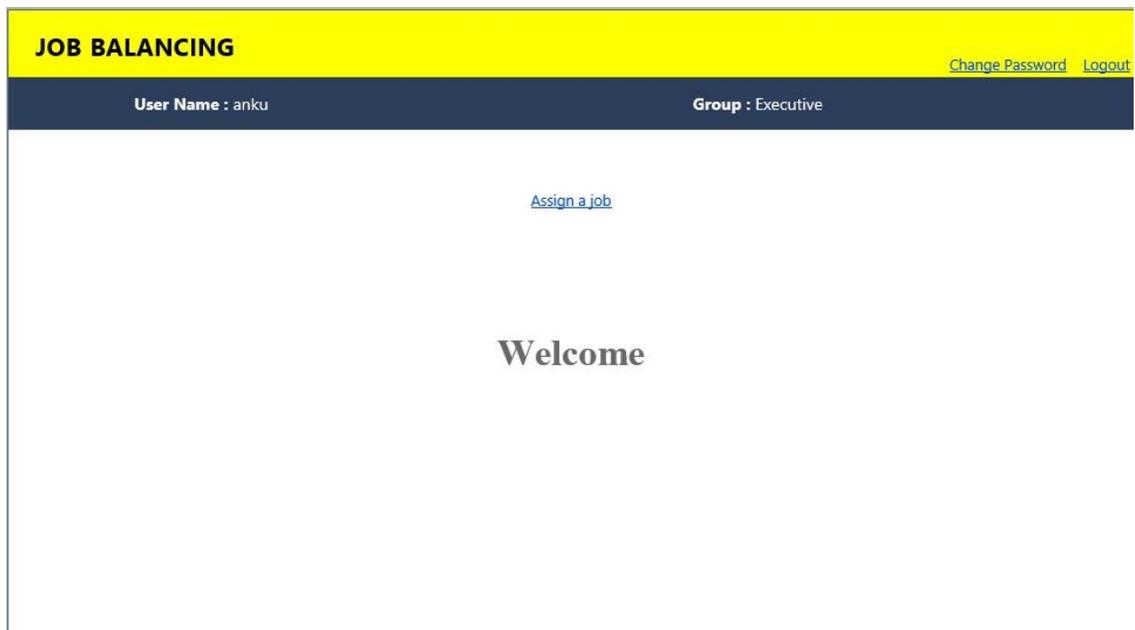

*Fig. 4: Lower Priority clients can only access the jobs*

But the question arises, how the priority is assigned if more than 1 client with same role want to access the job? It is done via FCFS. For e.g. if any 2 executives login to access the job, then the executive who login first is assigned the job and this whole process is managed by the Admin. It is shown in figure 5.

*Fig. 5 : FCFS +RBAC implemented according to priority*

### V. Conclusion and Future Scope:

By the implementation of RBAC and FCFS using .NET environment, the workload is reduced at a high cost. Moreover security is provided to all the clients working in an organisation. Although all these concepts are developed before, but a thorough implementation is proven in this paper using ASP.NET. Whole technique is proven to decrease the workload of the executer. But the workload can be reduced further if the RBAC is implemented using distributed environment instead of a single organisation. More work can be performed when one job execution depend on the execution of another job.

## References:

[1] P. Singh, S. Singh, "A New Advance Efficient RBAC to Enhance the Security in Cloud Computing", International *Journal of Advanced Research in Computer Science and Software Engineering, Volume 3, Issue 6*, June 2013

[2] A.Kaur, N.Bansal, " Cross Breed job scheduling for reducing server load using RBAC at Cloud", in International Journal of Advanced Research in Computer Science and Software Engineering, *Vol. 3*, pp. 330-333, May 2013

[3] Gitanjali, S. S. Sehra, J.Singh, "Policy Specification in Role based Access Control on Clouds", *International Journal of Computer Applications, Volume 75– No.1, August 2013*

[4] Y. Jung, M. Chung," Adaptive Security Management Model in the Cloud Computing Environment", *ICACT,* Feb. 7-10, 2010

[5] Z.Zhou, L. Wu, Z.Hong," Context-Aware Access Control Model for Cloud Computing",
*International Journal of Grid and Distributed Computing, Vol.6, No.6* (2013), pp.1-12

[6] Z.Tang, J. Wei, A.S.K. Li, R. Li," A New RBAC Based Access Control Model for Cloud Computing", *Springer-Verlag Berlin Heidelberg,* pp. 279–288, 2012.

[7] Reeja S L, "Role Based Access Control Mechanism In Cloud Computing Using Co – Operative Secondary Authorization Recycling Method", *Volume 2, Issue 10,* October 2012.

[8] Abdul Raouf Khan, "Access Control In Cloud Computing Environment", *Vol.7, No.5*, MAY 2012.